\documentclass[11pt,a4paper]{article}
\usepackage{jcappub}                    
\usepackage{amsmath,amssymb,mathtools}  
\usepackage{soul}                       
\usepackage[usenames,dvipsnames]{xcolor}
\usepackage{CJK}                        

\newcommand{\bb}{\mathbf}

\newcommand{\beq}{\begin{equation}}
\newcommand{\eeq}{\end{equation}}
\newcommand{\bed}{\begin{displaymath}}
\newcommand{\eed}{\end{displaymath}}
\def\bea{\begin{eqnarray}}
\def\eea{\end{eqnarray}}

\newcommand{\ba}{\begin{alignat}{3}}

\newcommand{\dd}{\text{d}}

\setstcolor{red}    
\setulcolor{red}    
\allowdisplaybreaks 

\title{Adiabatic regularisation of power spectra in $k$-inflation}

\author{Allan L. Alinea$^{a)}$,} 
\author{Takahiro Kubota$^{a), b), c)}$,}
\author{Yukari Nakanishi$^{a)}$}
\author{and Wade Naylor$^{a), d)}$}

\affiliation{ $a)$ Department of Physics, Osaka University, Toyonaka, Osaka 560-0043, Japan}
\affiliation{ $b)$ CELAS, Osaka University, Toyonaka, Osaka 560-0043, Japan}
\affiliation{ $c)$ Kavli IPMU (WPI), The University of Tokyo, 5-1-5 Kashiwa-no-Ha, Kashiwa City, Chiba 277-8583, Japan}
\affiliation{ $d)$ International College, Interdisciplinary Research Building, Osaka University, Toyonaka, Osaka 560-0043, Japan}

\emailAdd{alinea@het.phys.sci.osaka-u.ac.jp}
\emailAdd{kubota@celas.osaka-u.ac.jp}
\emailAdd{nakanishi@het.phys.sci.osaka-u.ac.jp}
\emailAdd{naylor@phys.sci.osaka-u.ac.jp}
 
\abstract{We look at the question posed by Parker {\it et al.} about the effect of UV regularisation on the power spectrum for inflation. Focusing on the slow-roll $k$-inflation, we show that up to second order in the Hubble and sound flow parameters, the adiabatic regularisation of such model leads to no difference in the power spectrum apart from certain cases that violate near scale-invariant power spectra. Furthermore, extending to non-minimal $k$-inflation, we establish the equivalence of the subtraction terms in the adiabatic regularisation of the power spectrum in Jordan and Einstein frames.}


\begin{document}
\begin{CJK*}{UTF8}{ipxm}    
\maketitle

\bigskip
\bigskip
\section{Introduction}
\label{intro}

Cosmological inflation \cite{Starobinsky:1979ty, Guth:1980zm, LiddlenLyth, Martin:2014} is nowadays considered a paradigm explaining the homogeneity and flatness problems associated with fine-tuning the initial conditions of the history of the Universe. What comes after solving these problems are the issues of convergence of theory and experiment, and also self-consistency of the theory. On the convergence side, with improved accuracy in the CMB measurements, inflationary predictions need to be more precise. Areas of consideration dealing with this involve calculating the power spectrum beyond the first slow-roll approximation \cite{Martin:2013}, loop corrections  \cite{Bartolo:2010bu}, etc. On the consistency side, the areas of consideration may involve non-gaussianity \cite{Maldacena:2002vr}, the tensor-to-scalar ratio \cite{Kinney:1998md}, and regularisation \cite{Parker:Toms}, among others. Owing to the absence of the explicit form of the Lagrangian describing inflation, there are indeed a lot of things to investigate. In this paper, on the side of consistency, we study the regularisation of the power spectrum in slow-roll $k$-inflation. This is effectively an extension of the earlier work by Urakawa and Starobinsky \cite{Urakawa:2009}.

The issue of regularisation of the power spectrum has been debated in recent years \cite{Parker:2007} where otherwise, a window or smearing function is introduced into the power spectrum to obtain physical results  \cite{Liddle:2000}. The need for regularisation may arise from the consideration of the two point function of cosmological perturbations. Consider for instance, the gauge-invariant (scalar) perturbation $\mathcal R(\eta, \mathbf x)$ \cite{Bardeen:1983} where $\eta$ is the conformal time. We have
\begin{equation}
    \big\langle
        \mathcal R(\eta, \bb x)\mathcal R(\eta, \bb y)
    \big\rangle
    = 
    \int \frac{\dd k}{k}
    \frac{\sin(k|\bb x-\bb y|)}{k|\bb x-\bb y|}
    \Delta^2_{\mathcal R}(k, \eta),
    \label{2point}
\end{equation}
where $\Delta^2_{\mathcal R}(k, \eta)$ is the dimensionless power spectrum, hereinafter referred to as simply \textit{power spectrum}, and is related to the Fourier transform $\mathcal R_k(\eta)$ by the definition
\begin{equation}
    \Delta^2_{\mathcal R}(k,\eta)
    \equiv
    \frac{k^3}{2\pi^2} |\mathcal R_k(\eta)|^2.
    \label{dimvar}
\end{equation}
There is a similar expression for the tensor perturbations. Given the Mukhanov-Sasaki equation (cf. (\ref{scalar})), the adiabatic condition tells us that $\Delta^2_{\mathcal R}(k,\eta)$ scales as $k^2$ in the UV limit. This implies that in the coincidence limit $\mathbf x\,\rightarrow\,\mathbf y$, the two-point function given above diverges. Such divergences are well known in quantum field theory with boundaries or on curved backgrounds, (e.g., see \cite{Fulling:Book}) and there is no dispute about this claim in the context of cosmology. What has been some issue of contention is the treatment of this divergence and its possible effect (if ever there would be) on the inflationary power spectrum for values of $k$ that are currently, observationally significant, as has been argued in \cite{Parker:2007}\footnote{Note that infrared (IR) divergences also appear in the massless de Sitter limit, but can be removed by using a scale factor that evolves from say, a radiation to a de Sitter phase, e.g., see \cite{Vilenkin:1982wt}.}.

One way to deal with the divergence of the two point function is to employ a method involving the subtraction of the divergent parts. Such a process of subtraction by virtue of (\ref{2point}), is correspondingly reflected on the power spectrum. In adiabatic regularisation \cite{Parker:Toms, Parker:1974qw, Fulling:1974pu, Bunch:1980vc}, the method we use in this work, a systematic way of subtracting mode-by-mode is used to regularised the power spectrum and end up with a finite two-point function in the coincidence limit. This is in some ways similar to the the brute force approach to regularising the Casimir effect by subtracting the mode sums for ``no" plates from those modes with plates \cite{Milton:Book}. For self-consistency, the subtraction terms have to be introduced not only for UV modes but for all modes. As a consequence, the ``frozen'' long-wavelength modes might as well be affected by the regularisation. As we shall see, under some assumptions, the original power spectrum that we simply call the ``bare'' power spectrum is not affected at all.

Recently, a detailed discussion by Bastero-Gil et al. \cite{Bastero-Gil:2013} has highlighted issues and reviewed previous work by other authors, including that of Parker and coworkers \cite{Parker:2007, Agullo:2009, Agullo:2009:II}. They reviewed alternate approaches \cite{Durrer:2009ii, Durrer:2011} that modified the adiabatic regularisation relevant to inflation by using a high momentum cutoff and it was argued \cite{Bastero-Gil:2013} that this approach is not appropriate as the cutoff modifies the standard energy equations for the fixed background equations. They went further and suggested that the variance associated with CMB temperature perturbations, $C_l$, depends on the temperature measured at two \textit{different} directions.\footnote{Essentially the same idea was emphasised in \cite{Finelli:2007fr} about six years earlier.} In other words, there is no need to consider the coincidence limit, $\mathbf x = \mathbf y$, in the power spectrum, cf. Eq. (\ref{2point}), and therefore the result is already finite.

In this article we take a pragmatic approach and assume that one {\it may} need to regularise the power spectrum. Looking at (\ref{2point}), the coincidence limit is a mathematical possibility that cannot simply be evaded by insisting that $\mathbf x \ne \mathbf y$ for some physical circumstances. Furthermore, when we consider interacting theories or for that matter the stress energy tensor \cite{Bunch:1980vc}, then it would still be necessary. As we shall see, and in the same vein as Urakawa and Starobinsky \cite{Urakawa:2009}, for theories with near scale-invariant power spectra (as imposed by experiment), the regularised power spectrum is equivalent to the bare one at late times. The subtraction terms ``wash out'' during inflation. An important point here is that we let the subtraction terms follow the time development, as in \cite{Urakawa:2009}, as opposed to the original idea of Parker \cite{Parker:2007}.

The outline of this paper is as follows. In the next section, Sec. \ref{model}, we lay down the set up for minimal and non-minimal $k$-inflation. Focusing on minimal $k$-inflation, we discuss in Sec. \ref{flow} the Mukhanov-Sasaki equation, express the effective potential in terms of the Hubble and sound flow parameters, and write down the expression for the bare power spectrum. Then in Sec. \ref{reg} we perform adiabatic regularisation of the power spectrum, analyse the behaviour of the subtraction term, and discuss its effect on regularised power spectrum. In Sec. \ref{conf} we go back to non-minimal $k$-inflation, establish the connection between the relevant quantities in the Jordan and Einstein frames, and demonstrate the equivalence of the subtraction terms in both frames. Finally, in Sec. \ref{conc}, we summarise our results and discuss the implication of regularisation in connection to the observed power spectrum.

\bigskip
\section{Set up: Minimal and Non-minimal \textbf{\textit{k}}-inflation}
\label{model}

\subsection{Minimal $k$-inflation}
In this work we are interested in the adiabatic regularisation of the power spectrum corresponding to a generalised single-field inflation termed as $k$-inflation, whose action for the inflaton field $\phi$ is an arbitrary function $P(X, \phi)$, where
\begin{align}
    X
    &\equiv
    -\frac{1}{2} \nabla^\mu\phi \nabla_\mu \phi.
    \label{actionMinimal}
\end{align} 
Denoting by $R$ the Ricci scalar, the action can be written as
\begin{align}
    S
    &=
    \frac{1}{2}
    \int \sqrt{-g}\, \dd^4 x
    \big[
        M_\text{Pl}^2\, R
        +
        2P(\phi,\, X)
    \big].
    \label{minimalkInflation}
\end{align}
For brevity, hereafter we take the Planck mass (squared) as unity; symbolically, $M_\text{Pl}^2 = 1$. When $P = X - V(\phi)$, where $V(\phi)$ is the inflaton potential, this action reduces to the canonical case. 

The conformally flat background spacetime corresponding to the action given above is described by the Friedmann-Lema\^itre-Robertson-Walker (FLRW) metric. For the perturbation about this background, we choose a gauge such that the inflaton fluctuation vanishes; \textit{i.e.}, $\delta \phi = 0$, and the metric can be written in terms of the ADM variables \cite{Arnowitt:1959ah} with the spatial part being specified by 
\begin{align}
    g_{ij}
    &=
    a^2(\eta)e^{2\mathcal R}\big(e^\gamma\big)_{ij},
    \qquad
    (\partial^i\gamma_{ij} = \gamma^{i}{}_i = 0).
\end{align}
Here, $\mathcal R$ is the curvature perturbation \cite{Bardeen:1983}, $\gamma_{ij}$ is the transverse and traceless tensor perturbation, $a$ is the scale factor, and $\eta$ is the conformal time defined in relation to the coordinate time $t$ as $\dd\eta \equiv \dd t/a$. Of the two perturbations, our main focus is on the curvature perturbation.

The background equations of motion, namely, the Friedmann and continuity equations read 
\begin{align}
    3{\mathcal H}^2
    &=
    a^2 E, 
    \nonumber
    \\[0.5em]
    {E^\prime}
    &=
    -3{\mathcal H}\left(E+P\right), 
    \label{friedmanContinuity}
\end{align}
where $E$ is the energy density and $\mathcal H$ is the conformal Hubble parameter defined by ${\mathcal H}\equiv{a^\prime}/a$ with the prime denoting differentiation with respect to the conformal time\footnote{We denote  differentiation with respect to the coordinate time $t$ by an overdot.}. In analogy with the continuity equation in the case of canonical single-field inflation, the quantity $P$ plays the role of ``pressure'' and we will call it as such from hereon. With $E$ being the 00-component of the energy momentum tensor $T^\mu{}_{\nu}$, the pressure is further related to the energy density by
\begin{align}
    E
    &=
    2XP_{,X}-P,
    \label{energy}
\end{align}
where $P_{,X}$ denotes the derivative of $P$ with respect to $X$. 

Upon decomposing the action with respect to $\mathcal R$ as $S = S^{(0)} + S^{(1)} + S^{(2)} + S^{(3)} + \cdots $, (see for instance, \cite{Chen:2010xka}) one arrives at the equation of motion from the second order action $S^{(2)}$.
\begin{align}
    v''_k 
    +
    \left(k^2 c_s^2 - \frac{z''}{z}\right)v_k
    =
    0.
    \label{scalar}
\end{align}
This is called the Mukhanov-Sasaki equation. In this equation, $v$ is related to the curvature perturbation by the definition $v \equiv z\mathcal R$, where $z^2 \equiv 2a^2 \epsilon/c_s^2 $, with $\epsilon$ being the (first) slow roll parameter defined by $\epsilon \equiv -\dot H/H^2$. The quantity $c_s$ defined as
\begin{align}
    c_s^2
    \equiv
    \frac{P_{,X}}{E_{,X}}
    =
    \frac{P_{,X}}{P_{,X}+2XP_{,XX}},
    \label{soundspeed}
\end{align}
is the ``speed of sound'' describing the speed of propagation of the scalar perturbations. In the last equation above, the right hand side follows from (\ref{energy}). The stable solutions corresponding to $c_s^2 > 0$ require that $P_{,X} > 0$ and $2X P_{,XX}+P_{,X} > 0$ for the Hamiltonian to be bounded from below and the field equations to be hyperbolic (see \cite{ArmendarizPicon:1999rj, Garriga:1999vw, Bruneton:2006gf} for more details).

\bigskip
\subsection{Non-minimal $k$-inflation}
\label{Sec : nonMinimalSetup}
The action given by (\ref{minimalkInflation}) can be further generalised to the case where the inflaton couples to the background geometry. In such a case, instead of the coefficient one, $R$ is multiplied by a function of the inflaton field, $f(\phi)$. The resulting action given by
\begin{align}
    S_\text{nm}
    &=
    \frac{1}{2}
    \int \sqrt{-g}\, \dd^4 x
    \big[
        f(\phi) R
        +
        2P(\phi,\, X)
    \big],
    \label{nonMinimalkInflation}
\end{align}
describes the extention of the \textit{minimal} $k$-inflation we have just have briefly discussed above to \textit{non-minimal} $k$-inflation. 

The introduction of $f(\phi)$ can lead to difficulties in the calculation of relevant physical quantities such as the $n$-point function of the scalar perturbations, power spectrum, spectral tilt, etc. Regularisation, the main subject of this work, is likewise affected prompting us to look for a simpler way to perform calculations. In search for such a simpler way, it would be better if we could carry over the set up above for the minimal case. One way to do it, and this is what we do here, is to transform the action above from the Jordan frame where it is written, to the Einstein frame. In the Einstein frame, the action takes the form of the minimal case (cf. (\ref{actionMinimal})):
\begin{align}
    S_\text{nm}
    &=
    \frac{1}{2}
    \int \sqrt{-\widehat g}\, \dd\widehat t\,\dd^3 x
    \big[
        \widehat R
        +
        2\widehat P(\phi,\, \widehat X)
    \big].
    \label{nonMinimalkInflationEinstein}
\end{align}
In Section \ref{conf}, we briefly review the main ideas covered in \cite{Kubota, Chiba:2008ia, Gong:2011qe} needed to arrive at the form of $S_\text{nm}$ given above and the relationship between the quantities with a hat in the Einstein frame and the quantities without a hat in the Jordan-frame. At this point, the important thing to note is that with the action written in the Einstein frame, all the equations written above for the minimal case are carried over to the non-minimal case with all the quantities involved replaced by their counterparts with a hat. It then remains for us to demonstrate the trick of going from the Jordan frame to the Einstein frame in the calculation of the subtraction terms needed to regularise the power spectrum in non-minimal $k$-inflation.

\bigskip
\bigskip
\section{The Bare Power Spectrum}
\label{flow}

The bare power spectrum involves the frozen modes $\mathcal R_k$ that exited the Hubble horizon during inflation. They are the long-wavelength solutions of the Einstein equations; more specifically, $|kc_s/aH| \ll 1$. The existence of constant mode solution in the long-wavelength limit for single-field inflation is guaranteed by the Einstein field equations (see \cite{Weinberg:2003sw} for details).

To derive the bare power spectrum, we start with the Mukhanov-Sasaki equation given by (\ref{scalar}). This equation resembles the equation of motion for a parametric oscillator. The product $kc_s$ corresponds to the angular frequency while $z''/z$ can be identified as the effective potential:
\begin{align}
    U_s(\eta)
    =
    \frac{z''}{z}.
\end{align}

The right hand side of the equation for $U_s$ can be written in terms of the Hubble flow parameters ($\epsilon_n$) and sound flow parameters ($\delta_n$) given by the recursive definitions \cite{Martin:2013}, \cite{Schwarz:2001vv}
\begin{align}
    \epsilon_{n+1}
    &\equiv
    \frac{\dd\ln{\epsilon_n}}{\dd N},
    \qquad
    \epsilon_0 
    \equiv 
    \frac{H_{i}}{H},
    \label{hubbleFlow}
    \\[0.5em]
    \delta_{n+1}
    &\equiv
    \frac{\dd\ln{\delta_n}}{\dd N},
    \qquad
    \delta_0 
    \equiv
    \frac{c_{s,i}}{c_s}.
    \label{soundFlow}
\end{align}
The quantity $N$ is the number of $e$-folds ($N \equiv \ln a/a_i$) and the subscript $i$ indicates an initial value. Note that with the above definition for $\epsilon_n$, $\epsilon_1$ is simply the first slow-roll parameter $\epsilon$ mentioned in the previous section. On the other hand, $\delta_1$ corresponds to the relative rate of change of the speed of sound. We can then rewrite the expression for the effective potential as \cite{Martin:2013}
\begin{align}
    U_s (\eta)
    =
    a^2 H^2 \left[2-\epsilon_1 +\frac{3}{2}\epsilon_2
    +
    \frac{1}{4}\epsilon_2^2 -\frac{1}{2}\epsilon_1\epsilon_2
    +
    \frac{1}{2}\epsilon_2 \epsilon_3 +(3-\epsilon_1 +\epsilon_2)\delta_1
    +
    \delta^2_1 +\delta_1 \delta_2 \right]~.
    \label{Us}
\end{align}

The effective potential as we can see, is expressed up to second order in terms of the Hubble and sound flow parameters. This \textit{exact} form of $U_s$ brought about by the introduction of the Hubble and sound flow parameters, makes the Mukhanov-Sasaki equation above amenable to semi-analytic approximations such as the uniform approximation \cite{Olver}, allowing the calculation of the (bare) power spectrum to a high level of precision with respect to $\epsilon_n$ and $\delta_n$ evaluated at horizon crossing or the so-called turning point \cite{Martin:2013}. Up to second order with respect to the Hubble and sound flow parameters, the bare power spectrum can be written as \cite{Zhu:2014wfa} 
\begin{align}
    \Delta_{\mathcal R}^{2\,(b)}
    &=
    c_0\frac{\bar H^2}{8\pi^2\bar \epsilon_1 \bar c_s}
    \bigg[
        1 + \left(\frac{429}{181} - \ln 2\right)\bar \delta_1
        +
        \left(\ln 4 - \frac{496}{181}\right)\bar \epsilon_1
        +
        \left(\ln 2 - \frac{67}{181}\right)\bar \epsilon_2
        \nonumber
        \\[0.5em]
        &\qquad
        +\,
        \left(\frac{2095}{1086} + \frac{\ln^2 2}{2} 
        - 
        \frac{600\ln 8}{1267}\right)\bar \delta_1^2
        + 
        \left(\frac{4865}{1629} - \frac{\pi^2}{24} + \frac{\ln^2 2}{2} 
            - \frac{429\,\ln 2}{181}
        \right)\bar \delta_1 \bar \delta_2
        \nonumber
        \\[0.5em]
        &\qquad
        +\,
        \left(\frac{811\,\ln 2}{181} - \frac{541}{181} - 2\ln^2 2\right)
            \bar \delta_1 \bar \epsilon_1
        +
        \left(\frac{293}{181} + 2\ln^2 2 
            - 
            \frac{315\ln 4}{181}
        \right)\bar \epsilon_1^2
        \nonumber
        \\[0.5em]
        &\qquad
        +\,
        \left(-\frac{56}{181} - \ln^2 2+ \frac{315\ln 2}{181}\right) 
        \bar \delta_1 \bar \epsilon_2
        +
        \left(
            \frac{\pi^2}{12} - \frac{4231}{1629} + \ln^2 2 
            + \frac{47\ln 2}{181}
        \right)
        \bar \epsilon_1 \bar\epsilon_2
        \nonumber
        \\[0.5em]
        &\qquad
        +
        \left(-\frac{11}{362} + \frac{\ln^2 2}{2} 
            - \frac{67\ln 2}{181}
        \right)\bar \epsilon_2^2
        +
        \left(\frac{\pi^2}{24} 
            - \frac{86}{1629} - \frac{\ln^2 2}{2} 
            + \frac{67\ln 2}{181}
        \right)
        \bar \epsilon_1\bar \epsilon_3
    \bigg],
\end{align}
where the over bar indicates evaluation at the turning point,
\begin{align}
    k\bar c_s
    &=
    -\frac{\bar\nu}{\bar\eta},
    \qquad
    \text{with}
    \qquad
    \nu^2 \equiv \tfrac{1}{4} + \eta^2 U_s(\eta),
\end{align}
and $c_0 = 181/9e^3 \simeq 1.0013$. For our purposes, we simply ignore the numerical error in $c_0$ and take it as exactly unity. Evaluated at horizon crossing, $k c_{s*} = a_* H_*$, the bare power spectrum can be cast in the form
\begin{align}
    \Delta_{\mathcal{R}}^{2\,(b)}
    &=
    \frac{H_*^2 }{8\pi^2 \epsilon_{1*}c_{s*}}
    (1 + \alpha_{1*} + \alpha_{2*}),
    \label{barePowerSpec}
\end{align}
where the label `*' denotes evaluation at horizon crossing and $\alpha_{1*}$ and $\alpha_{2*}$ are first and second order respectively, in terms of $\epsilon_{(n\le 3)*}$ and $\delta_{(n\le 2)*}$. They are given by \cite{Zhu:2014wfa} 
\begin{align}
    \alpha_{1*}
    &=
    \left( \frac{429}{181} - \ln 3 \right)\delta _{1*}
    +
    \left(\ln 9 - \frac{496}{181}\right)\epsilon _{1*}
    +
    \left(\ln 3 - \frac{67}{181}\right)\epsilon _{2*}
    \nonumber
    \\[0.5em]    
    \alpha_{2*}
    &=
    \left(
	\frac{457}{362} + \frac{\ln^2 3}{2}
	-
	\frac{248\ln 3}{181} - \frac{64\ln 2}{1267}
    \right)\delta _{1*}^2
    +
    \left(
    	\frac{517}{543} + 2\ln^2 3 - \frac{630\ln 3}{181}
    \right)\epsilon _{1*}^2
    \nonumber
    \\[0.5em]
    &\qquad
    +\,
    \left(
    	\frac{\pi ^2}{12} - \frac{3688}{1629} 
    	+
    	\ln^2 3 + \frac{47\ln 3}{181}
    \right)\epsilon _{1*}\epsilon _{2*}
    +
    \left(
    	\frac{329}{1086} + \frac{\ln^2 3}{2}
    	-
    	\frac{67\ln 3}{181}
    \right)\epsilon _{2*}^2
    \nonumber
    \\[0.5em]
    &\qquad
    +\,
    \left(
    	-\frac{\pi ^2}{24} + \frac{4865}{1629}
    	+
    	\frac{\ln^2 3}{2} - \frac{429\ln 3}{181}
    \right)\delta _{1*}\delta _{2*}
    +
    \left(
    	-\frac{718}{543} - 2\ln^2 3 + \frac{811\ln 3}{181}
    \right)\delta _{1*}\epsilon _{1*}
    \nonumber
    \\[0.5em]
    &\qquad
    +\,
    \left(
    	\frac{13}{543} - \ln^2 3 + \frac{315\ln 3}{181}
    \right)\delta _{1*} \epsilon _{2*}
    +
    \left(
    	\frac{\pi ^2}{24} - \frac{86}{1629}
    	-
    	\frac{\ln^2 3}{2} + \frac{67\ln 3}{181}
    \right)\epsilon _{2*}\epsilon _{3*}.
\end{align}
We will use the expression for the bare power spectrum above later in the calculation of the regularised power spectrum.

\bigskip
\bigskip
\section{Adiabatic Regularisation of the Power Spectrum}
\label{reg}

\subsection{Derivation of the Subtraction Terms}
Let $\Delta^{2\,(b)}_{\mathcal R}$ and $\Delta^{2\,(s)}_{\mathcal R}$ be the bare power spectrum and the corresponding subtraction term respectively. The regularised power spectrum $\Delta^{2\,(r)}_{\mathcal{R}}$ is then simply the difference of the two:
\begin{align}
    \Delta^{2\,(r)}_{\mathcal R}
    &=
    \Delta^{2\,(b)}_{\mathcal R} - \Delta^{2\,(s)}_{\mathcal R} ~.
    \label{regPowerSpec}
\end{align}
The calculation of the bare power spectrum through the relation $\Delta^{2\,(b)}_{\mathcal R} \,\propto\, |v^{(b)}_k|^2/z^2$ involves solving the Mukhanov-Sasaki equation for $v^{(b)}_k$. Likewise, the calculation of the subtraction term $\Delta^{2\,(s)}_{\mathcal R}$, our main objective in this subsection, involves solving $v^{(s)}_k$ that satisfies the Mukhanov-Sasaki equation to a certain desired adiabatic order. In adiabatic regularisation, one assumes the \textit{adiabatic condition} that 
\begin{align}
    v^{(s)}_k
    \sim
    \frac{1}{\sqrt{2\omega_k(\eta)}}
    e^{-i\int^{\eta} \dd \eta'\,\omega_k(\eta')},
    \qquad
    \left(\omega_k^2 = k^2 c_s^2\right)
\end{align}
to lowest adiabatic order and considers the ansatz
\begin{align}
    v_k^{(s)}(\eta)
    \sim
    \frac{1}{\sqrt{2W_k(\eta)}}\,
    e^{-i\int^\eta \dd \eta'\, W_k(\eta')}.
    \label{ansatzV}
\end{align}
Ommitting the subscript $k$ for brevity, the quantity $W$ satisfies the homogeneous differential equation given by \cite{Parker:Toms}
\begin{align}
    W^2 - \frac{3}{4}\left(\frac{W'}{W}\right)^2
    +
    \frac{1}{2} \frac{W''}{W}
    -
    \Omega^2
    &=
    0,
    \label{workingEq}
\end{align}
with $\Omega^2 \equiv k^2 c_s^2 - U_s(\eta)$, reminiscent of the Mukhanov-Sasaki equation, and the effective potential $U_s$ is given by (\ref{Us}). The process of solving (\ref{workingEq}) involves expanding $W$ into a series of terms $\omega^{(n)}$ where the non-negative integer $n$ denotes the adiabatic order. Symbolically, one has
\begin{align}
    W = \omega^{(0)} + \omega^{(1)} + \omega^{(2)} + \cdots,
\end{align}
and (\ref{workingEq}) is solved order-by-order. As we can see, in adiabatic regularisation, the calculation of $v^{(s)}_k$ and hence, of the subtraction term for the power spectrum, boils down to the computation of $\omega^{(n)}$. 

It is rather tedious but straightforward to calculate $\omega^{(n)}$ through the differential equation (\ref{workingEq}). In this work, following the minimal subtraction scheme\footnote{See \cite{Finelli:2007fr} for a different point of view about using the minimal subtraction scheme.}, we go up to second adiabatic order:
\begin{align}
    \omega^{(0)}
    &=
    kc_s,
    \qquad
    \omega^{(1)}
    =
    0,
    \nonumber
    \\[0.5em]
    \omega^{(2)}
    &=
    -\frac{(aH)^2}{kc_s}\big(
        1 + \delta\varepsilon + \delta c_s
    \big).
\end{align}
In the last equation above,
\begin{align}
    \delta\epsilon 
    &\equiv 
    \frac{1}{2}\left(
        -\epsilon_1 +\frac{3}{2}\epsilon_2 
        +
        \frac{1}{4}\epsilon_2^2 
        -
        \frac{1}{2}\epsilon_1\epsilon_2 
        +
        \frac{1}{2}\epsilon_2 \epsilon_3
    \right)
    \nonumber
    \\[0.5em]
    \delta c_s 
    &\equiv
    \frac{1}{8}\,\delta_1 \big(
        10 - 2\epsilon_1 + 4\epsilon_2 + 3\delta_1 +2 \delta_2
    \big).
\end{align}
Now, since $v_k^{(s)}(\eta)$ is as given by the ansatz above namely (\ref{ansatzV}), we see that we need to only invert $W = \omega^{(0)} + \omega^{(2)} + \cdots$, to finally calculate $\Delta^{2\,(s)}_{\mathcal{R}}$. It follows that in the large-scale limit, specifically, when $|kc_s/aH| \ll 1$,
\begin{align}
    \Delta^{2\,(s)}_{\mathcal R}
    &=
    \frac{k^3}{2\pi^2} \left|\frac{v_k}{z}\right|^2
    =
    \frac{k^2}{4\pi^2 z^2 c_s}
    \left[
        1 + (1 + \delta\epsilon + \delta c_s)\left(
            \frac{aH}{kc_s}
        \right)^2
    \right] ~,
    \nonumber
    \\[0.5em]
    \Delta^{2\,(s)}_{\mathcal R}
    &=
    \frac{H^2}{8\pi^2 \epsilon_1c_s}
    \left(1 + \delta\epsilon + \delta c_s\right),
    \label{subPowerSpec}
\end{align}
The last equation above is our sought-for equation for the subtraction term in this subsection. 

\bigskip
\bigskip
\subsection{Behaviour of the Subtraction terms and the Power Spectrum}
In this subsection, we analyse the behaviour of the subtraction term and its effect on the power spectrum. The regularised power spectrum is given by (\ref{regPowerSpec}). With the expressions for $\Delta^{2\,(b)}_{\mathcal{R}}$ and $\Delta^{2\,(s)}_{\mathcal{R}}$ given by (\ref{barePowerSpec}) and the last of (\ref{subPowerSpec}) respectively, in hand, we find
\begin{align}
    \Delta^{2\,(r)}_{\mathcal R}
    &=
    \frac{H_*^2}{8\pi^2 \epsilon_{1*}c_{s*}}
    \bigg[
        1 + \alpha_{1*} + \alpha_{2*}
        - 
        (1 + \delta\epsilon + \delta c_s)
        \left(\frac{H}{H_*}\right)^2
        \left(\frac{\epsilon_{1*}}{\epsilon_1}\right)
        \left(\frac{c_{s*}}{c_s}\right)
    \bigg].
    \label{genRegPowerSpec}
\end{align}
This expression involves Hubble and sound flow parameters up to \textit{second} order. The presence of $c_s$ reminds us of the more general $k$-inflation that includes the single-field canonical inflation as a special case. The expression above for $\Delta^{2\,(r)}_{\mathcal R}$ generalises the result in \cite{Urakawa:2009} involving the canonical single-field inflation. Indeed, when $c_s = 1$, up to \textit{first} order in terms of the Hubble flow parameters, we have\footnote{We ignore small numerical error in the expression for the bare power spectrum.} 
\begin{align}
    \Delta^{2\,(r)}_{\mathcal R}\bigg|_{c_s = 1, \mathcal O(\epsilon)}
    =
    \frac{H_*^2}{8\pi^2 \epsilon_{1*}}
    \bigg[
        1 
        + 
        \delta\varepsilon_{*}
        -
        (1 + \delta\varepsilon)
        \left(\frac{H}{H_*}\right)^2
        \left(\frac{\epsilon_{1*}}{\epsilon_1}\right)        
    \bigg],
    \label{regPowSpecial}
\end{align}
where 
\begin{align}
    \delta\varepsilon_{*}
    &=
    (2\epsilon_{1*} + \epsilon_{2*})(2 - \ln 2 - \gamma_\text{E}) 
    - 
    2\epsilon_{1*},
    \nonumber
    \\[0.5em]
    \delta\varepsilon
    &=
    -\frac{1}{2}\epsilon_1
    +
    \frac{3}{4}\epsilon_2,
\end{align}
with $\gamma_\text{E}$ being the Euler-Mascheroni constant. Equation (\ref{regPowSpecial}) is the same\footnote{The expression for the regularised power spectrum in \cite{Urakawa:2009} is effectively \textit{first} order with respect to the Hubble flow parameters even though the subtraction term is expressed up to second order in Hubble flow parameters because the bare power spectrum is calculated only up to first order.} equation for the regularised power spectrum derived in \cite{Urakawa:2009}.

Going back to the general expression for $\Delta^{2\,(r)}_{\mathcal R}$ given by (\ref{genRegPowerSpec}), we focus on the last term inside the pair of square brackets. For the second factor involving the Hubble parameter, we see from the definitions of the first slow roll parameter ($\epsilon_1 = \epsilon \equiv -\dot H/H^2$) and the number of $e$-folds ($\dd N \equiv \dd\ln a$) that
\begin{align}
    \left(\frac{H}{H_*}\right)^2
    &=
    \exp\left(
        -2\int_{N_*}^{N}\dd N'\, \epsilon_1(N')
    \right).
\end{align}
For the third factor involving the first slow roll parameter (or the first Hubble flow-parameter), we find from the definition of $\dot \epsilon_1$ given by (\ref{hubbleFlow}) that
\begin{align}
    \left(
        \frac{\epsilon_1}{\epsilon_{1*}}
    \right)
    &=
    \exp\left(
        \int_{N*}^N \dd N'\,\epsilon_2(N')
    \right).
\end{align}
It follows that in accord with the result in \cite{Urakawa:2009}, when the speed of sound is constant, the subtraction term is suppressed by a factor of
\begin{align}
    \left(\frac{H}{H_*}\right)^2
    \left(
        \frac{\epsilon_{1*}}{\epsilon_{1}}
    \right)
    &=
    \exp\left(
        -\int_{N_*}^{N}\dd N'\, (2\epsilon_1 + \epsilon_2)
    \right),
    \label{hEpsilon}
\end{align}
and the bare power spectrum at horizon exit matches the regularised power spectrum at late times. 

The fourth factor in the second term inside the pair of square brackets in the equation for $\Delta^{2\,(r)}_{\mathcal{R}}$ involves the speed of sound. In the simplest possible scenario where $c_s^2 \ne \text{constant}$, symmetry in the form of scale invariance requires that the relative change of the speed of sound measured by $\epsilon_s$ defined as\footnote{Needless to say, $\epsilon_s$ is nothing but the negative of the first sound flow parameter $\delta_1$. We choose to also use the notation in \cite{Khoury:2008wj} for emphasis.}
\begin{align}
    \epsilon_s
    \equiv
    \frac{1}{H}
    \frac{\dot c_s}{c_s},
    \label{defEpsilonS}
\end{align}
be constant and ``entangled'' with the first slow-roll parameter. In particular\footnote{As pointed out in \cite{Khoury:2008wj} there is another case wherein the power spectrum is scale-invariant namely, $\epsilon_s = \frac{2}{5}(3 - 2\epsilon_1)$, corresponding to a contracting universe with growing speed of sound. In this work, we do not deal with contracting universe.}, $\epsilon_s = -2\epsilon_1 = \text{constant}$, corresponding to an expanding background with decreasing speed of sound \cite{Khoury:2008wj}. We elevate this relation to the more general case where both $\epsilon_1$ and $\epsilon_s$ are functions of $\eta$ and add a small correction $\beta$ describing deviation from that required by scale invariance. Symbolically, we take 
\begin{align}
    \epsilon_s(\eta)
    =
    -2\epsilon_1(\eta) + \beta(\eta).
    \label{soundCorrection}
\end{align}
Note that as shown in \cite{Khoury:2008wj}, when $\epsilon_1,\, \epsilon_s \ll 1$, that is, when we have a quasi-de Sitter background and nearly constant speed of sound, the equation above for $\epsilon_s$ leads to the usual slow roll result for the spectral tilt derived in \cite{Garriga:1999vw} namely, $n_s - 1 \approx -2\epsilon_1 - \epsilon_s - \epsilon_2$.

The last equation above for $\epsilon_s$ together with its definition given by (\ref{defEpsilonS}) yields
\begin{align}
    \frac{c_{s*}}{c_s}
    =
    \exp\left(
        2\int_{N_*}^N \dd N'\, \epsilon_1(N')
    \right)
    \exp\left(
        -\int_{N_*}^N \dd N'\, \beta(N')
    \right).
    \label{csBehave}    
\end{align}
Upon multiplying (\ref{hEpsilon}) by (\ref{csBehave}), the exponential term involving $\epsilon_1$ in the expression for $c_{s*}/c_s$ given above is cancelled. The regularised power spectrum then becomes
\begin{align}
    \Delta^{2\,(r)}_{\mathcal R}
    &=
    \frac{H_*^2}{8\pi^2 \epsilon_{1*}c_{s*}}
    \left[
        1 + \alpha_{1*} + \alpha_{2*}
        - 
        (1 + \delta\epsilon + \delta c_s)
        \exp\left(
            -\int_{N_*}^N \dd N'\, (\epsilon_2 + \beta)
        \right)        
    \right]
    \label{powerSpecFinal}
\end{align}
For an expanding universe, if on the average if not monotonically, $\epsilon_1$ increases with $N$ then $\epsilon_2 = \dd\ln\epsilon_1/\dd N$ is positive. With $\epsilon_2$ being a dominant term in the argument of the exponential function above, it follows that the factor $(1 + \delta\epsilon + \delta c_s)$ is exponentially suppressed. As the product $(\frac{H^2}{H_*})^2(\frac{\epsilon_{1*}}{\epsilon_1})(\frac{c_{s*}}{c_s})$ decays with $N$ as the universe expands, the regularised power spectrum given by (\ref{powerSpecFinal}) tends to the bare power spectrum. 

Note that as pointed out above, scale invariance ties $\epsilon_s$ and $\epsilon_1$ together as $\epsilon_s = -2\epsilon_1 = \text{constant}$. The constant in the right hand side is not necessarily zero. When the speed of sound is constant, this special relation from which we based (\ref{soundCorrection}) reduces to the trivial case where $\epsilon_1 = 0$. If we insist that only $\epsilon_s$ can vanish independent of $\epsilon_1$, the quantity $\beta$ cannot just be a small correction about $\epsilon_s = -2\epsilon_1$ which seems misleading if we are to stick near scale invariance in the general case. This tells us that when the special case where $c_s^2 = \text{constant}$ is considered, meaning that we are dealing with the canonical single-field inflation, one has to go back to the more general relation given by (\ref{genRegPowerSpec}) and take the limit as $c_s^2\,\rightarrow\,c_{s*}^2$, instead of using (\ref{powerSpecFinal}). However, this turns (\ref{powerSpecFinal}) and (\ref{regPowSpecial}) somewhat into two disjoint relations. A resolution would be to have $|\beta| < \delta_b$, where $\delta_b$ is nonnegative and second order in the sound and Hubble flow parameters, so as to have a small deviation about the prescription $\epsilon_s = -2\epsilon_1$, but allow for the flexibility for (\ref{powerSpecFinal}) to converge to (\ref{regPowSpecial}) in the limit that $\epsilon_s\,\rightarrow\, 0$. In other words, we enlarge the range of $\beta$ to $-\delta_b <\beta < 2\epsilon_1$ to encompass the result of \cite{Urakawa:2009}. With this new range, we have the same behaviour of the last term inside the pair of square brackets in (\ref{powerSpecFinal}); that is, $(1 + \delta\epsilon + \delta c_s)$ is exponentially suppressed by a decaying factor.

\bigskip
\bigskip
\section{Frame Independence of the Subtraction Terms}
\label{conf}
Given the more complicated structure of the non-minimal $k$-inflation action, it is useful to employ a conformal transformation to go to the Einstein frame. The equivalence of the \textit{bare} power spectra of non-minimal $k$-inflation for curvature perturbations in Jordan and Einstein frames was explicitly shown in \cite{Kubota} (see also e.g., \cite{Makino} for a detailed discussion of the case where $P = X - V$ but $\phi$ couples non-minimally to $R$). In relation to this, the issue of equivalence for the \textit{regularised} power spectra should also be checked. With this goal in mind we briefly review the conformal properties of non-minimal $k$-inflation.

\par The FLRW \textit{background} metric in the Einstein frame is given by 
\begin{align}
    \dd\widehat{s}\,^2
    =
    \Omega^{2}\, \dd s^{2}
    =
    -d\widehat{t}\,^2
    +
    \widehat{a}^2(\,\widehat{t}\;)\delta_{ij}\dd x^i \dd x^j,
    \label{nmet}
\end{align}
where $\dd \widehat t = \Omega\,\dd t$ and the conformal scale factor is related to $a(t)$ by $\widehat a(\,\widehat{t}\;) = \Omega \big(\phi(t) \big) \, a(t)$. Note that $\widehat{a}(\,\widehat{t}\;)$ does not contain a fluctuation part. On the other hand, the \textit{perturbed} metric takes the form in accord with our chosen gauge mentioned in Sec. \ref{flow}, that is, $\delta\phi = 0$ and the metric can be written in ADM decomposition as
\begin{align}
    \dd \widehat s\,^2
    &=
    -\widehat N^2 \,\dd\,\widehat t\,^2
    +
    \widehat h_{ij}\big(
        \dd x^i + \widehat N^i\,\dd \widehat t\,
    \big)\big(
        \dd x^j + \widehat N^j\,\dd \widehat t\,
    \big),
\end{align}
with the spatial component being given by 
\begin{align}
    \widehat h_{ij}
    =
    \widehat{a}^2(\,\widehat{t}\;) e^{\widehat {\mathcal R}} 
    \big(e^{\widehat \gamma}\big)_{ij},
    \qquad
    \big(
        \partial^i \widehat \gamma_{ij}
        =
        \widehat\gamma^i{}_i 
        =
        0
    \big)\,.
\end{align}

\bigskip
With the metric in hand, we can then proceed to rewrite the action in Jordan frame to that in the Einstein frame. In the Jordan frame, 
\begin{align}
    S_\text{nm}
    &=
    \frac{1}{2} \int \dd t\dd^3\vec x\,
    \sqrt{-g} \left[
        f(\phi) R + 2P(\phi,\,X)
    \right].
\end{align}
Using the conformal transformation properties of the Ricci scalar \cite{Fujii:2003pa},\cite{Wald:1984rg},
\begin{align}
    R
    =
    \Omega^{2}\Big [
        \widehat{R} + 6\widehat{\square} \ln \Omega 
        -
        6\widehat{g}^{\mu\nu}
        (\partial  _\mu\ln\Omega) (\partial _\nu\ln\Omega )
    \Big]~,
    \label{eq:ricchi1}
\end{align}
the action can be written as 
\begin{align}
    S_\text{nm}
    &= 
    \frac 1 2\int \dd\widehat{t}\: \dd^{3}\vec{x}
    \:\frac{\sqrt{-\widehat{g}}}{\Omega^4}
    \Big\{\Omega^2 f(\phi)
        \Big[ \widehat R + 6\widehat{\square} \ln \Omega 
            -
            6 \widehat{g}\,^{\mu\nu} (\partial _\mu\ln\Omega) 
            (\partial _\nu\ln\Omega )
        \Big]
    + 
    2P \Big\}.
    \label{eq:ricchi2}
\end{align}
In order to arrive at the Einstein frame by making the coefficient 
of the scalar curvature unity, we set $\Omega (\phi) = \sqrt{f(\phi)} $. Furthermore, we define
\begin{align}
    \widehat{X} 
    &\equiv 
    -\frac{1}{2}\widehat{g}\,^{\mu \nu}
    \partial _{\mu} \phi 
    \partial _{\nu} \phi 
    =\frac{1}{\Omega^{2}} X~,
    \nonumber
    \\[0.5em]
    \widehat{P}(\phi,\widehat X)
    &\equiv
    \frac{1}{\Omega^{4}} P(\phi, X) 
    -
    3\widehat{g}\,^{\mu\nu} (\partial _\mu\ln\Omega) 
    (\partial _\nu\ln\Omega )
    \nonumber 
    \\[0.5em]
    &=
    \frac{1}{\Omega^{4}} P(\phi, X) 
    + 
    \frac{6\widehat{X}}{\Omega ^{2}}
    \left (\frac{\dd\Omega}{\dd\phi}\right )^{2}~.
    \label{eq:phatprelation}
\end{align}
After integration by parts to eliminate $\widehat \square$, the action becomes \cite{Kubota}
\begin{align}
    S_\text{nm}
    &=
    \frac 1 2\int \dd\widehat{t} \: \dd^{3}\vec{x}\:
    \sqrt{-\widehat{g}}\Bigl[ 
    \widehat{R} + 2 \widehat{P}(\phi,\widehat X) \Bigr].
    \label{einsteinMod}
\end{align}
This is our desired expression for the action in the Einstein frame---the one we advertised in Sec. \ref{Sec : nonMinimalSetup}.

Let us now focus on the scalar perturbations and their two point correlation functions. We know that the free vacuum and the ``time-developed vacuum" in the interaction picture are identical in both Einstein and Jordan frames \cite{Kubota}. Moreover, the curvature perturbations are also identical ($\widehat {\mathcal R} = \mathcal R$) \cite{Chiba:2008ia, Gong:2011qe}. It follows that the equal-time bare cosmological correlation functions and the bare power spectra in each frame are the same; symbolically,
\begin{align}
    \big|
        \widehat{\mathcal{R}}^{(b)}_k (\eta) 
    \big|^{2}
    &=
    \big|
        \mathcal{R}^{(b)}_k (\eta) 
    \big|^{2},
    \nonumber
    \\[0.5em]
    \widehat \Delta^{2(b)}_{\mathcal R}(k,\eta)
    &=
    \Delta^{2(b)}_{\mathcal R}(k,\eta),
\end{align}
where the label `($b$)' indicates bare and the conformal time in the left hand side for both equations above is written without a hat because $\widehat \eta = \eta$. Furthermore, we may expect for physical reasons, that the equal-time regularised correlation functions computed in each frame agree with each other. This implies that the corresponding subtraction terms for the power spectrum are equal. Symbolically, we may expect
\begin{align}
    \big|\widehat{\mathcal{R}}^{(s)}_k (\eta)\big|^{2}
    &=
    \big|\mathcal{R}^{(s)}_k (\eta)\big|^{2},
    \nonumber
    \\[0.5em]
    \widehat\Delta^{2(s)}_{\mathcal R}(k,\eta)
    &=
    \Delta^{2(s)}_{\mathcal R}(k,\eta),    
    \label{equalSubtract}
\end{align}
where the label `($s$)' corresponds to subtraction term.

That (\ref{equalSubtract}) holds at least in adiabatic regularisation can be shown through the following argument. The action for non-minimal $k$-inflation can be written in the Einstein frame (resembling the form of the minimal case) as we have just discussed above. Such an action just like that of the minimal case can be expanded with respect to $\widehat{\mathcal R}$ as $S_\text{nm} = S^{(0)}_\text{nm} + S^{(1)}_\text{nm} + S^{(2)}_\text{nm} + S^{(3)}_\text{nm} + \cdots$. Term by term, $S^{(n)}_\text{nm}$ taken from the action in Jordan frame is equal to $S^{(n)}_\text{nm}$ taken from the action in Einstein frame where $n \ge 0$ (see \cite{Kubota} for an explicit calculation for $n \le 3$ ). Furthermore, the equations of motion derived from $S^{(2)}_\text{nm}$ in Jordan and Einstein frames are equivalent. In adiabatic regularisation, the curvature perturbation corresponding to the subtraction term satisfies the Mukhanov-Sasaki equation to a certain desired adiabatic order. Since $\widehat {\mathcal R}^{(s)}_k$ and ${\mathcal R}^{(s)}_k$ satisfy equivalent Mukhanov-Sasaki equations (one in the Einstein  and one in the Jordan frame) to the same certain desired adiabatic order and obey the same boundary conditions, then they are one and the same and (\ref{equalSubtract}) holds.

Let us demonstrate this explicitly at least up to second adiabatic order. From the second order action in the Einstein frame we arrive at the Mukhanov-Sasaki equation just like that in the minimal case except that the quantities involved have a hat:
\begin{align}
    \widehat v''_k 
    +
    \left(
        \widehat c_s\!^2 k^2 - \frac{\widehat z\,''}{\widehat z}
    \right)\widehat v_k
    &=
    0,
    \nonumber
    \\[0.5em]
    \Leftrightarrow
    \qquad
    \left[
        \frac{\dd}{\dd\widehat{t}}\left(
            \widehat{a}\,^3
            \frac{\widehat{\epsilon}_1}{{\widehat c_s}\!^2}
            \frac{\dd}{\dd\widehat{t}}
        \right)
        +
        \widehat{a}\,\widehat{\epsilon}_1 k^2
    \right]
    \widehat{\mathcal{R}}_k (\eta)
    &=
    0.    
    \label{scalarEinsteinFrame}
\end{align}
The quantities with a hat such as $\widehat{z}$ are defined in terms of quantities in the Einstein frame:
\begin{align}
    \widehat {\mathcal R}
    \equiv
    \frac{\widehat v}{\widehat z},
    \quad
    \widehat{z}
    \equiv
    \frac{\sqrt{2\widehat{\epsilon}_1}\,\widehat{a}}{\widehat{c}_s},
    \quad
    {\widehat{c}_s}\!^2
    \equiv
    \frac{\widehat{P}_{,\widehat{X}}}{\widehat{E}_{,\widehat{X}}}
    =
    \frac{\widehat{P}_{,\widehat{X}}}{
        \widehat{P}_{,\widehat{X}}
        +
        2\widehat{X}\widehat{P}_{,\widehat{X}\widehat{X}}
    },
    \quad
    \widehat \epsilon_1
    =
    -\frac{1}{\widehat H^2}
    \frac{\dd \widehat H}{\dd\widehat t},
    \quad
    \widehat a 
    =
    \Omega a.
    \label{eq:Ezcs}
\end{align}
We can calculate $|\widehat {\mathcal R}^{(s)}_k|^2$ for non-minimal $k$-inflation in the Einstein frame in the same way as that in the minimal coupling model.
\begin{align}
    \big|\widehat{\mathcal{R}}^{(s)}_k (\eta)\big|^2
    &=
    \frac{1}{2k \widehat z\,^2 \widehat c_s}
    \bigg[
        1 + (1 + \delta\widehat \epsilon + \delta \widehat c_s)\bigg(
            \frac{\widehat a \widehat H}{k\widehat c_s}
        \bigg)^2
    \bigg]
    \nonumber
    \\[0.5em]
    \big|\widehat{\mathcal{R}}^{(s)}_k (\eta)\big|^2
    &=
    \frac{1}{2\widehat{z}\,^2 \widehat{c}_s k}
    \left[
        1 + \frac{1}{2{\widehat{c}_s}\!^2k^2}
        \frac{\widehat{z}\,''}{\widehat{z}}
        +
        \frac{1}{\widehat{c}_s\!^2 k^2}
        \left(
            \frac{1}{4}\frac{{\widehat{c}_s}\!''}{\widehat{c}_s}
            -
            \frac{3}{8}\frac{{\widehat c_s}\!'^2}{{\widehat c_s}\!^2}
        \right)
    \right].
    \label{eq:Esub}
\end{align}

On the other hand, the equation that the scalar perturbation obeys in the Jordan frame was derived in \cite{Qiu:2011}. The authors used the second-order action involving $\mathcal R$ (which is $\zeta$ in their notation,) and arrived at the equation of motion that we rewrite using our own notation as
\begin{equation}
    u''_k +\left(c_{s,\text{eff}}^2 k^2 
    - 
    \frac{z''_{\text{eff}}}{z_{\text{eff}}}\right) u_k 
    =
    0,
    \label{jordanMukhanovSasaki}
\end{equation}
where $u_k \equiv \mathcal R_k/z_\text{eff}$ and $c_{s,\text{eff}}$ can be regarded as the effective sound speed. With appropriate normalisation,
\begin{align}
    z_{\text{eff}}^2
    &=
    6e^{2\theta}\left(\frac{H}{\dot{\theta}} - 1\right)^2 
    +
    \frac{2a^2 \Sigma}{\dot{\theta}^2},
    \nonumber
    \\[0.5em]
    c_{s,\text{eff}}^2
    &=
    -\frac{2}{z^2_{\text{eff}}}e^{2\theta}\bigg(
        \frac{H}{\dot{\theta}} - 1 + 
        \frac{\ddot{\theta}}{\dot{\theta}^2}
    \bigg),
    \label{effCsz}
\end{align}
where $\theta\equiv\frac{1}{2}\ln{fa^2} = \ln \,\Omega a $ and $\Sigma\equiv XP_{,X}+2X^2 P_{,XX}$. Using (\ref{jordanMukhanovSasaki}), we can calculate $|\widehat {\mathcal R}^{(s)}_k|^2$ as before:
\begin{equation}
    |\mathcal{R}^{(s)}_k (\eta)|^2
    =
    \frac{1}{2z_{\text{eff}}^2 c_{s,\text{eff}} k}
    \bigg\{
        1 + \frac{1}{2c_{s,\text{eff}}^2 k^2}
        \frac{z''_{\text{eff}}}{z_{\text{eff}}}
        +
        \frac{1}{c_{s,\text{eff}}^2 k^2}
        \bigg[
            \frac{1}{4}\frac{c''_{s,\text{eff}}}{c_{s,\text{eff}}}
            -
            \frac{3}{8}\bigg(
                \frac{c'_{s,\text{eff}}{}}{c_{s,\text{eff}}}
            \bigg)^2 
        \bigg]
    \bigg\}~.
\end{equation}
Note that this has the same form as $|\widehat {\mathcal R}^{(s)}_k|^2$  (in the Einstein frame) given by the second of (\ref{eq:Esub}).

It remains for us to establish the relationship between $\{u_k,\, z_\text{eff},\, c_{s,\text{eff}}\}$ and $\{\widehat v_k,\, \widehat z,\, \widehat c_{s}\}$. To express $\widehat{z}$ and $\widehat{c}_s\!^2$ in terms of $\theta,\, H,$ and $\Sigma$, use the definitions given by (\ref{eq:Ezcs}). Moreover, define $\widehat{\Sigma} \equiv \widehat{X}\widehat{P}_{,\widehat{X}} + 2\widehat{X}^2 \widehat{P}_{,\widehat{X}\widehat{X}}$ and use the Friedmann equation in the Einstein frame. We have
\begin{align}
    \widehat{c}_s\!^2 
    =
    \frac{\widehat{\epsilon}_1 \widehat{H}^2}{\widehat{\Sigma}},
    \qquad
    \widehat{z}\,^2 
    =
    \frac{2\widehat{\epsilon}_1     
    \widehat{a}^2}{\widehat{c}_s^2}
    =
    \frac{2\widehat{a}^2 \widehat{\Sigma}}{\widehat{H}^2},
    \label{eq:Ecs}
\end{align}
where \cite{Kubota},
\begin{align}
    \widehat{\Sigma}
    &=
    e^{-4\theta}a^4\left[
        \Sigma 
        + 
        3e^{2\theta}a^{-2}\left(
            \dot{\theta}-H
        \right)^2
    \right],
    \nonumber
    \\[0.5em]
    \widehat{H}
    &=
    e^{-\theta}a\dot{\theta},
    \qquad
    \widehat{\epsilon}_1 
    =
    -\left(
        \frac{H}{\dot{\theta}} - 1 
        + 
        \frac{\ddot{\theta}}{\dot{\theta}^2}
    \right).
\end{align}
Upon performing substitution into (\ref{eq:Ecs}) we find
\begin{align}
    \widehat{z}\,^2
    &=
    \big(e^{\theta}\big)^2\dot{\theta}^2
    e^{-4\theta}a^2\bigg[
        6e^{2\theta}\left(\frac{H}{\dot{\theta}} - 1\right)^2
        +
        \frac{2a^2 \Sigma}{\dot{\theta}^2}
    \bigg]
    \big(e^{-\theta}a\dot{\theta}\big)^{-2}
    \\[0.5em]
    \widehat{c}_s\!^2 
    &=
    -\bigg(
        \frac{H}{\dot{\theta}} - 1 
        + 
        \frac{\ddot{\theta}}{\dot{\theta}^2}
    \bigg)
    \cdot 
    2e^{2\theta}\bigg[
        6e^{2\theta}\left(\frac{H}{\dot{\theta}} - 1\right)^2
        +
        \frac{2a^2 \Sigma}{\dot{\theta}^2}
    \bigg]^{-1}
    \nonumber
    \\[0.5em]
    &=
    -\frac{2}{z^2_{\text{eff}}}e^{2\theta}\left(
        \frac{H}{\dot{\theta}} - 1 
        + 
        \frac{\ddot{\theta}}{\dot{\theta}^2}
    \right)
\end{align}
It follows from (\ref{effCsz}) that $\widehat c_s = c_{s,\text{eff}},\, \widehat z = z_{\text{eff}}$, and we can identify $u_k = \widehat v_k$. This implies that $\widehat {\mathcal R}_k^{(s)}(\eta) = {\mathcal R}_k^{(s)}(\eta)$, which in turn, means that the subtraction terms in both the Einstein and Jordan frame are identical at least up to second adiabatic order; that is, $  \widehat\Delta^{2(s)}_{\mathcal R}(k,\eta) = \Delta^{2(s)}_{\mathcal R}(k,\eta)$. Hence, we get the same result for the regularised power spectrum.

It is enough that we have demonstrated the equivalence of the subtraction terms in the Jordan and Einstein frames. Our aim is simply to put forward a simple way to carry out adiabatic regularisation for the non-minimal coupling case by performing the calculation in the Einstein frame. In the minimal case, we went beyond this to look into the behaviour of the subtraction term under some simplifying assumptions. Owing to the complicated structure and wide possibilities involved in the non-minimal coupling case, we leave the analysis of the behaviour of the subtraction term and the regularised power spectrum for future studies.

\bigskip
\bigskip
\section{Concluding Remarks}
\label{conc}

In this article we have extended the work of Urakawa and Starobinsky \cite{Urakawa:2009} to the case where the speed of sound is varying. In particular, we have studied the adiabatic regularisation of the power spectrum in minimal slow-roll $k$-inflation. We found out that up to second order in the sound and Hubble flow parameters, the adiabatic regularisation of such model leads to no difference in the bare power spectrum apart from certain cases that violate near scale invariant power spectra. In such cases, the relative speed of sound may vary well beyond the condition of scale invariance namely, $\epsilon_s = -2\epsilon_1 = \text{constant}$.

In perspective, by following the subtraction terms long enough after horizon crossing, we ended up with simply the bare power spectrum. One starts with quantum perturbations corresponding to the bare power spectrum and the subtraction terms that are also quantum in nature. After the first horizon crossing, the former freezes and becomes indistinguishable from classical perturbations while the latter decays. As the continuing expansion of the Universe turns the modes to the frozen superhorizon modes, it also washes out the subtraction terms; the regularised power spectrum tends to the bare power spectrum. This tells us that UV regulators mainly affect small scale quantum fluctuations. The scalar modes stretched by inflation well beyond the Hubble horizon that gave birth to the large-scale structures such as galaxies and clusters of galaxies that we observe today, possibly carry no imprint of such a UV regularisation.

We have also explored the adiabatic regularisation of the power spectrum for non-minimal $k$-inflation. The extension from the minimal to the non-minimal case poses complications due to the coupling of the inflaton field with the background geometry. Inspired by earlier work \cite{Kubota} on the equivalence of the calculations of the bare power spectrum in the Jordan and Einstein frames, we looked into the possibility of extending that equivalence for the subtraction terms obtained via adiabatic regularisation. In this work, we have shown the formal equivalence of the subtraction terms in the Jordan and Einstein frames; hence, we can choose to work in the Einstein frame where calculations are carried out just like for that of the minimal case.

Looking ahead, this work connects to many areas of concern that are worthy of future investigation. The mechanism of classicalisation for instance, is still under active consideration; eg., see \cite{Starobinsky:1994, Martin:2005, Martin:2012, Martin:2012-2, Burgess:2014, Lim:2014}. It is tempting to ponder that the ``washing out" of subtraction terms during inflation might be related to classicalisation. Furthermore, for free theories, like what we have considered here, the adiabatic regularisation of the energy-momentum tensor and two-point function has been shown to be equivalent to a renormalisation of the action, e.g., see \cite{Bunch:1980vc, Haro:2010}; however, this is unclear for interacting theories. Hence it would be interesting to look at interacting theories, the accompanying loop corrections, and the possible interplay of adiabatic regularisation and other renormalisation/regularisation schemes for the calculation of the physical power spectrum (see for instance, \cite{Markkanen:2013}, for a preliminary study along this line but dealing with the energy-momentum tensor). In the case of matter interactions we could also look at how adiabatic regularisation affects the reheating process for the generation of matter and the reheating temperature. As we mentioned, inflation tends to ``wash out" the subtraction terms, but this may not be the case for matter fields coupled to the inflaton during reheating,  post-inflation.

\acknowledgments

We would like to thank members of the High Energy Theory group, Osaka University; in particular we have benefited from useful discussions with Nobuhiko Misumi, Kin-ya Oda, and Yutaka Hosotani.


\end{CJK*}
\end{document}